\documentstyle[prl,aps,twocolumn,epsf]{revtex}
\input{psfig}
\def\break#1{\pagebreak \vspace*{#1}}
\begin{document}


\title{Riccati parameter modes from Newtonian free damping motion by 
supersymmetry}

\author{Haret C. Rosu 
and Marco A. Reyes
}

\address{ 
Instituto de F\'{\i}sica de la Universidad de Guanajuato, Apdo Postal
E-143, Le\'on, Guanajuato, M\'exico 
 }

\maketitle
\widetext

\begin{abstract}

We determine the class of damped modes $\tilde{y}$ which are
related to the common free damping modes $y$ by supersymmetry. They are
obtained by employing the
factorization of Newton's differential equation of motion for the free
damped oscillator by means of the general solution of the corresponding
Riccati equation together with Witten's method of constructing the
supersymmetric partner operator.
This procedure leads to one-parameter families of
(transient) modes for each of the three types of free damping,
corresponding to a particular type of
antirestoring acceleration
(adding up to the usual Hooke restoring acceleration) of
the form $a(t)=\frac{2\gamma ^2}{(\gamma t+1)^{2}}\tilde{y}$, where $\gamma$
is
the family parameter that has been chosen as the inverse of the Riccati
integration constant. In supersymmetric terms, they represent all those one
Riccati parameter damping modes having the same Newtonian free damping partner
mode.

\end{abstract}
\vskip 0.1in

PACS number(s):  03.20.+i
\vskip 0.1in


\narrowtext

The damped oscillator (DO) is a cornerstone of physics and a primary
textbook example in classical mechanics. Schemes of analogies allow
its extension to many areas of physics where the same basic concepts
occur with merely a change in the meaning of the symbols. Apparently,
there might hardly be anything new to say about such an obvious case.
However, in the following we would like to exhibit a different and
nice feature of damping resulting from the mathematical procedure of
factorization of its differential equation. In the past, the factorization
of the DO differential equation (Newton's law) has been tackled by a few
authors \cite{do} but not in the framework that will be presented herein.
Namely, recalling that such factorizations are common
tools in Witten's supersymmetric quantum mechanics \cite{W} and imply
particular solutions of Riccati equations known as superpotentials,
we would like to explore here the factoring of the DO
equation by means of the general solution of the Riccati equation, a
procedure that has been used in physics by Mielnik \cite{M} for the
quantum harmonic oscillator. In other words, our goal here is to exploit the
nonuniqueness of the factorization of second-order differential operators,
on the example of the classical damped oscillator. By doing this one may hope 
to gain insight into the free damping motion.
We write the ordinary DO Newton's law in the form
$$
Ny\equiv
\left(\frac{d^2}{dt^2}+2\beta\frac{d}{dt}+\beta ^2\right)y=
(\beta ^2-\omega _{0}^{2})y=\alpha ^2 y~,
\eqno(1)
$$
i.e., we already added a $\beta ^2y$ term in both sides in order to perform
the factoring. The coefficient $2\beta$ is the friction constant per unit
mass and
$\omega _{0}$ is the natural frequency of the oscillator. The factorization
$$
\left(\frac{d}{dt}+\beta\right)\left(\frac{d}{dt}+\beta\right)y=\alpha ^{2}y
\eqno(2)
$$
follows, and previous authors \cite{do} discussed the classical cases of
underdamping ($\alpha ^2<0$),
critical damping ($\alpha ^2=0$), and
overdamping ($\alpha ^2>0$)
\break{1.77in}
in terms of the first-order differential equation
$$
Ly\equiv
\left(\frac{d}{dt}+\beta \right)y_{\pm}=\pm \alpha y_{\pm}~.
\eqno(3)
$$
It follows that $y_{\pm}=e^{-\beta t\pm \alpha t}$ and one can build
through their superposition the general solution
as $y=e^{-\beta t}(Ae^{\alpha t}+Be^{-\alpha t})$. Thus,
for free underdamping, the general solution can be written as $y_{u}=
\tilde{A}e^{-\beta t}\cos (\sqrt{-\alpha ^2}t +\phi)$, where $\tilde{A}
=2\sqrt{|AB|}$ and $\phi={\rm Arcos}(\frac{A+B}{\tilde{A}})$, whereas the
overdamped general solution is
$\tilde{A}e^{-\beta t}{\rm cosh} (\alpha t +\phi)$, where $\tilde{A}
=2\sqrt{|AB|}$ and $\phi={\rm Arcosh}(\frac{A+B}{\tilde{A}})$. The critical
case is special but well known \cite{do}, having the general solution of
the type $y_{c}=e^{-\beta t}(A+Bt)$~.

Since, as we mentioned, the factorization given by Eq. (2) may not be the
only one possible, let us now write the more general factorization
$$
N_{g}y\equiv
\left(\frac{d}{dt}+f(t)\right)\left(\frac{d}{dt}+g(t)\right)y=\alpha ^{2}y~,
\eqno(4)
$$
where $f(t)$ and $g(t)$ are two functions of time.
The condition that $N_{g}$ be identical to $N$ leads to
$f(t)+g(t)=2\beta$ and $g^{'}+fg=\beta$, that can be combined in the
following Riccati equation
$$
-f^{'}-f^{2}+2\beta f=\beta ^2~.
\eqno(5)
$$
By inspection, one can easily see that a first solution to this equation is
$f(t)=\beta$ ($g(t)=\beta$), which is the common case discussed by all the
previous authors \cite{do}. Changing the dependent variable
to $h(t)=f(t)-\beta$, we get
a simpler form of the Riccati equation, i.e., $h^{'}(t)+h^2=0$, with the
particular solution $h(t)=0$. However, the general solution is
$h(t)=\frac{1}{t+T}=\frac{\gamma}{\gamma t+1}$, as one can easily check.
The constant of integration
$T=1/\gamma$ occurs as a new time scale in the problem; see below.
Therefore, there is the more general factorization of the DO equation than
Eq.~(2) 
$$
A^{+}A^{-}y\equiv
\left(\frac{d}{dt}+\beta +\frac{\gamma}{\gamma t+1}\right)
\left(\frac{d}{dt}+\beta-\frac{\gamma}{\gamma t+1}\right)y
=\alpha ^{2}y~.
\eqno(6)
$$
A few remarks are in order. While the linear operator $L=\frac{d}{dt}+\beta$
has $y_{\pm}$ as eigenfunctions with eigenvalues $\pm \alpha$, the quadratic
operator $N$ has $y_{\pm}$ as degenerate eigenfunctions, with the same
eigenvalue $\alpha ^{2}$. On the other hand, the new linear operators $A^{+}$
and $A^{-}$ do not have $y_{\pm}$ as eigenfunctions since
$A^{+}y_{\pm}=(\pm \alpha +\frac{\gamma}{\gamma t+1})y_{\pm}$ and
$A^{-}y_{\pm}=(\pm \alpha -\frac{\gamma}{\gamma t+1})y_{\pm}$, although
the quadratic operator $N_{g}=A^{+}A^{-}$ still has $y_{\pm}$ as degenerate
eigenfunctions at eigenvalue $\alpha ^{2}$.
We now construct, according to the ideas of supersymmetric quantum
mechanics \cite{W},
the supersymmetric partner of $N_{g}$
$$
\tilde{N}_{g}=A^{-}A^{+}=\frac{d^2}{dt^2}+2\beta\frac{d}{dt}+\beta ^{2}-
\frac{2\gamma ^2}{(\gamma t+1)^2}~.
\eqno(7)
$$
This second-order damping operator contains the additional last term with
respect to its initial partner, which, roughly speaking, is the
Darboux transform term \cite{D} of the
quadratic operator. The important property of this operator is the
following. If $y_{0}$ is an eigenfunction of $N_{g}$, then $A^{-}y_{0}$
is an eigenfunction of $\tilde{N}_{g}$ since $\tilde{N}_{g}A^{-}y_{0}=
A^{-}A^{+}A^{-}y_{0}=A^{-}N_{g}y_{0}$ and $N_{g}y_{0}=\alpha ^2y_{0}$,
implying $\tilde{N}_{g}(A^{-}y_{0})=A^{-}N_{g}y_{0}=\alpha ^2
(A^{-}y_{0})$. The conclusion is that $\tilde{N}_{g}$ has the same type
of ``spectrum" as $N_{g}$, and therefore as $N$.
The eigenfunctions $\tilde{y}_{\pm}$
can be constructed if one knows the eigenfunctions $y_{\pm}$ as 
$$
\tilde{y}_{\pm}=A^{-}y_{\pm}=\left(\frac{d}{dt}+\beta -
\frac{\gamma}{\gamma t+1}\right)
y_{\pm}
\eqno(8)
$$
and thus
$$
\tilde{y}_{\pm}=\left(\pm \alpha -\frac{\gamma}{\gamma t+1}\right)
e^{-\beta t\pm \alpha t}.
\eqno(9)
$$
These modes make up a one-parameter family of damping
eigenfunctions that we interpret as follows.
We write down the usual form of the Newton law corresponding to the
Newton operator $\tilde{N}_{g}$,
$$
\left(\frac{d^2}{dt^2}+2\beta\frac{d}{dt}+\omega _{0}^{2}-
\frac{2\gamma ^2}{(\gamma t+1)^2}\right)\tilde{y}=0~.
\eqno(10)
$$
Examination of this law
shows that the term $2\gamma ^2/(\gamma t+1)^2\tilde{y}$
can be interpreted as a time-dependent antirestoring acceleration
(because of the
minus sign in front of it) producing in the transient period
$t\leq 1/\beta$ the damping modes given by $\tilde{y}$ above.

We present now separately the $\tilde{y}$  families of modes calculated as
superpositions of the modes $\tilde{y} _{\pm}$ for the three types of
free damping.

(i) For underdamping $\beta ^{2}<\omega _{0}^{2}$, let
$\alpha =i\omega _1$, where $\omega _1=\sqrt{\omega _{0}^{2}-\beta ^2}$.
The original eigenfunction is
$y_{u}=\tilde{A}_{u}\cos(\omega _{1}t+\phi)e^{-\beta t}$,
while the supersymmetric family
is $\tilde{y} _{u}= -\tilde{A} _{u}
[\omega _1\sin(\omega _1t+\phi)+\frac{\gamma}{\gamma t+1}
\cos(\omega _1t+\phi)]e^{-\beta t}$.

(ii) In the case of critical damping $\beta ^2=\omega _{0}^{2}$, the general
free solution is
$y_{c}=Ae^{-\beta t}+Bte^{-\beta t}$, whereas the tilde solution will
be $\tilde{y}_{c}=[\frac{-A\gamma}{\gamma t+1}+\frac{D}{\gamma ^2}
(\gamma t +1)^2]e^{-\beta t}$.
There is a difficulty in this case since
$\tilde{y}_{+}=A^{-}y_{+}=\frac{-A\gamma}{\gamma t+1}e^{-\beta t}$,
whereas $\tilde{y}_{-}=A^{-}y_{-}=\frac{B}{\gamma t+1}e^{-\beta t}\propto
\tilde{y}_{+}$. To find the independent $\tilde{y}_{-}$ solution we write
$\tilde{y}_{-}=z(t)\tilde{y}_{+}$ and determine the function $z(t)$ from
$\tilde{N}_{g}\tilde{y}_{-}=0$. The result is $z(t)=
\frac{C(\gamma t+1)^3}{\gamma ^3}$, where $C$ is an arbitrary constant,
and therefore $\tilde{y}_{-}=D\frac{(\gamma t+1)^2}{\gamma ^2}e^{-\beta t}$,
$D$ being another arbitrary constant.

(iii) For overdamping $\beta ^2>\omega _{0}^{2}$, the initial free
general solution is $y_{o}=\tilde{A} _{o}e^{-\beta t}\cosh(\alpha t+\phi)$,
whereas the $\gamma$ solution is
$\tilde{y} _{0}=-\tilde{A} _{o}e^{-\beta t}[\alpha \sinh(\alpha t+\phi)-
\frac{\gamma}{\gamma t +1}\cosh (\alpha t+\phi)]$.

Plots corresponding to these cases are presented in Figs. 1-3.
We note that in the limit $\gamma \rightarrow 0$ the modes 
$\tilde{y}_{\omega,\beta,\gamma}$ are going to the Newtonian damping modes
$y_{\omega,\beta}$ for all three classes of free damping motion.
Moreover, we placed ourselves herein in the well-behaved regime of motion,
i.e., for time and parameter ranges where the modes do not grow with time and
their amplitudes are finite.
However, from the point of view of the $\gamma$ parameter the modes $\tilde{y}$
are always singular, i.e., they blow up
at some negative time moment for positive $\gamma$ and at some positive instant 
for negative $\gamma$. Such blow-up solutions are quite well known in nonlinear
physics. On the other hand, even the Newtonian modes
$y_{\omega,\beta}$ are growing with time in the past or for negative $\beta$ in
the 
future (divergent and flutter instabilities are textbook knowledge \cite{tb}). 
What we claim here is that when one starts a damping-type measurement after
a ``mechanical" blow-up phenomenon, Riccati parameter modes may be present.
As we said, they may also occur before a blow-up phenomenon (for negative
$\gamma$), an equally important case. In this situation
the Riccati parameter distinguishes them from more common instability modes. 
Thus, an extended, Riccati-type parametrization of free damping can indeed be useful. 
The complexification of the Riccati parameter adds one more parameter to the 
Riccati damping modes. Depending on the sign of the imaginary part, new 
contributions to either damping or destabilization of the modes occur.

In summary, what we have obtained here are Riccati parameter families of 
damping modes related to the Newtonian free damping ones by means of Witten's 
supersymmetric scheme and the general Riccati solution.

\section*{Acknowledgment}
This work was partially supported by CONACyT Project 4868-E9406.



\newpage
\begin{figure}[htb]
 \begin{minipage}[t]{7.5cm}
 \vspace{-2.2cm}
 \hspace*{-.8cm}
 \psfig{figure=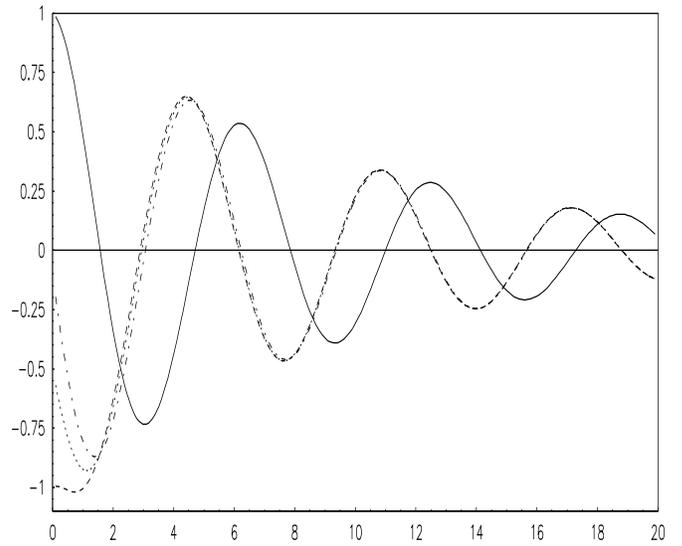,height=19cm,width=10.1cm}
 \end{minipage}
 \vspace*{-9.2cm}
\caption{Initial free underdamped mode of the type
$y_{u}=e^{-t/10}\cos t$ (bold curve)
and members of its $\gamma$ family of
supersymmetric damping modes $\tilde{y} _{u}=-e^{-t/10}(\sin t +
\frac{\gamma}{\gamma t+1}\cos t)$ 
for the following values of parameter $\gamma$: dashed curve - 1;
bold dashed curve - 1/2; solid curve - 1/10.
}
 \vspace*{-3mm}
\label{fig1}
\end{figure}

\newpage

\begin{figure}[htb]
 \begin{minipage}[t]{7.5cm}
 \vspace{-2.2cm}
 \hspace*{-.7cm}
 \psfig{figure=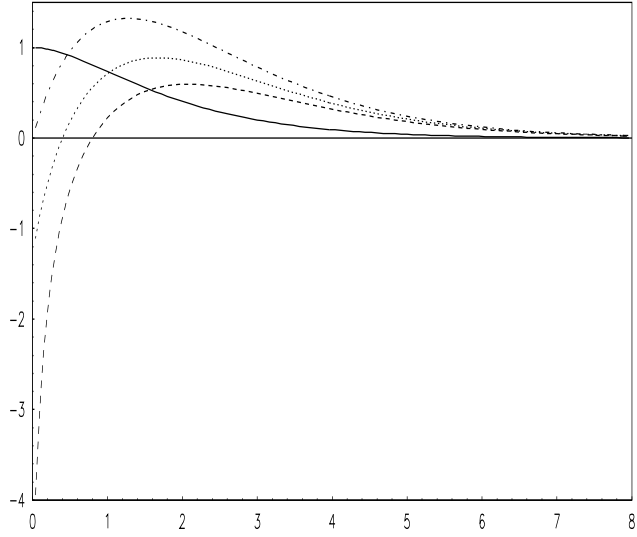,height=19cm,width=10cm}
 \end{minipage}
 \vspace*{-9.2cm}
\caption{Initial free critical damping mode $y_{c}=e^{-t}(1+t)$ (bold curve)
and members of the corresponding $\gamma$ 
family $\tilde{y} _{c}=e^{-t}
(\frac{-\gamma}{\gamma t+1}+\frac{(\gamma t+1)^2}{\gamma ^2})$
 for the $\gamma$ parameter
taking the following values: dashed curve - 5; bold-dashed curve - 5/3;
dot-dashed curve - 1.}
 \vspace*{-3mm}
\label{fig2}
\end{figure}

\newpage

\begin{figure}[htb]
 \begin{minipage}[t]{7.5cm}
 \vspace{-2.2cm}
 \hspace*{-.7cm}
 \psfig{figure=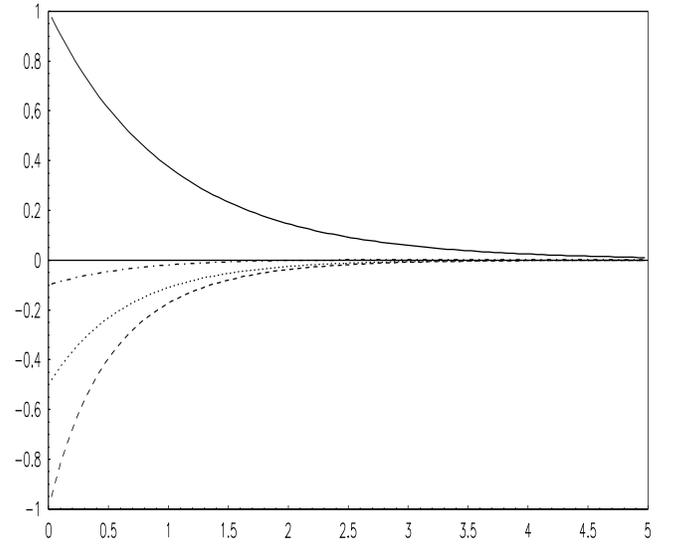,height=19cm,width=10cm}
 \end{minipage}
 \vspace*{-9.2cm}
\caption{Initial free overdamped mode of the type $y_{o}=e^{-t}\cosh (t/5)$
and members of its supersymmetric $\gamma$ family
$\tilde{y} _{o}=e^{-t}[\frac{1}{5}\sinh (t/5)-
\frac{\gamma}{\gamma t +1}\cosh(t/5)]$ for the following values of the parameter $\gamma$:
dashed curve - 1; bold dashed curve - 1/2; solid curve - 1/10.}
 \vspace*{-3mm}
\label{fig3}
\end{figure}

\newpage

\begin{figure}[htb]
 \begin{minipage}[t]{7.5cm}
 \vspace{-2.2cm}
 \hspace*{3cm}
 \psfig{figure=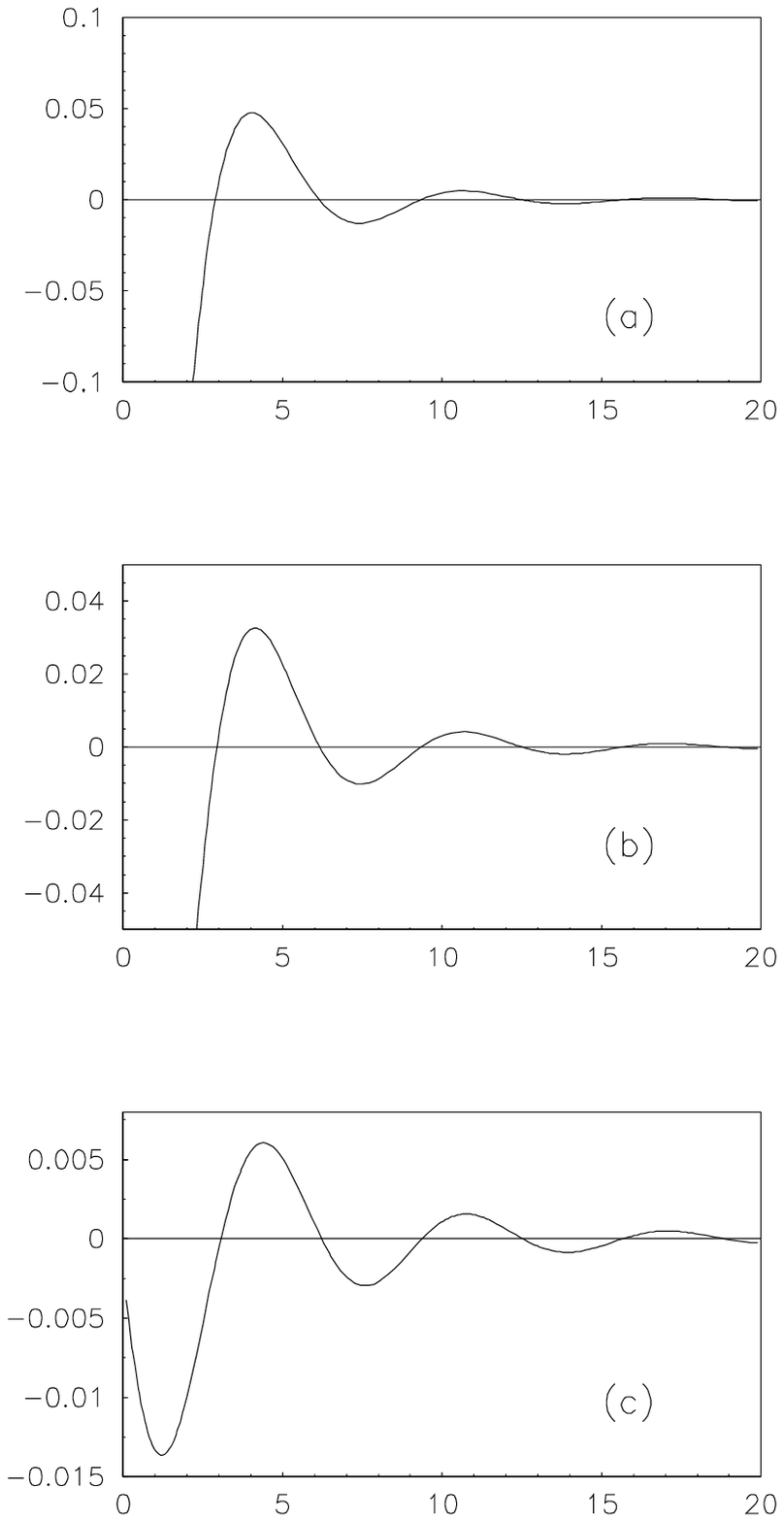,height=19cm,width=10cm}
 \end{minipage}
 \vspace*{-1cm}
\caption{The antirestoring acceleration for the underdamped modes 
$\tilde{y} _{u}$ at the 
values of the $\gamma$ parameter: (a) 1; (b) 1/2; (c) 1/10. }
 \vspace*{-3mm}
\label{fig4}
\end{figure}

\newpage

\begin{figure}[htb]
 \begin{minipage}[t]{7.5cm}
 \vspace{-2.2cm}
 \hspace*{3cm}
 \psfig{figure=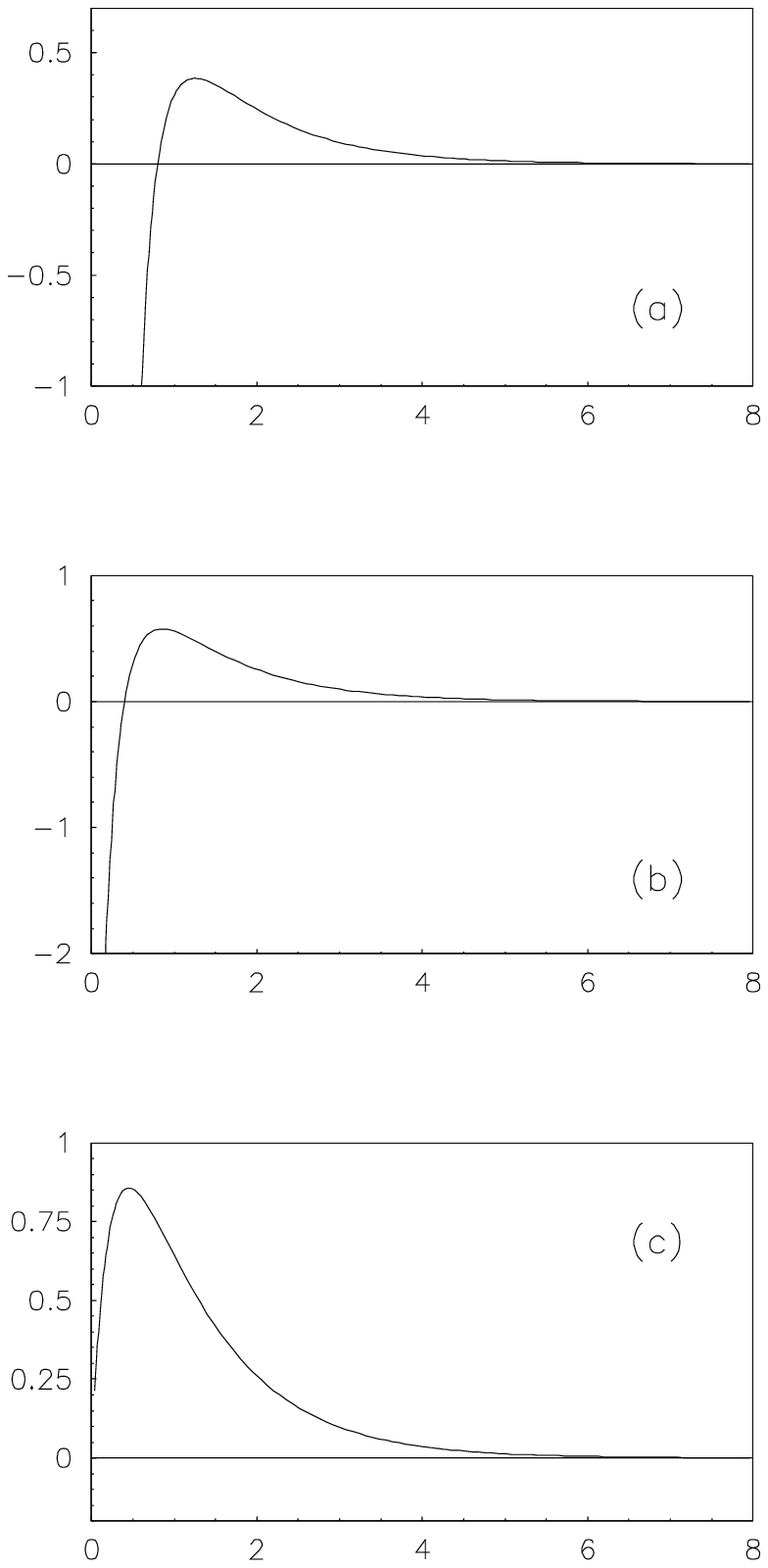,height=19cm,width=10cm}
 \end{minipage}
 \vspace*{-1cm}
\caption{The antirestoring acceleration for the critical modes 
$\tilde{y} _{c}$ at the values
of the $\gamma$ parameter: (a) 5; (b) 5/3; (c) 1.}
 \vspace*{-3mm}
\label{fig5}
\end{figure}

\newpage

\begin{figure}[htb]
 \begin{minipage}[t]{7.5cm}
 \vspace{-2.2cm}
 \hspace*{2cm}
 \psfig{figure=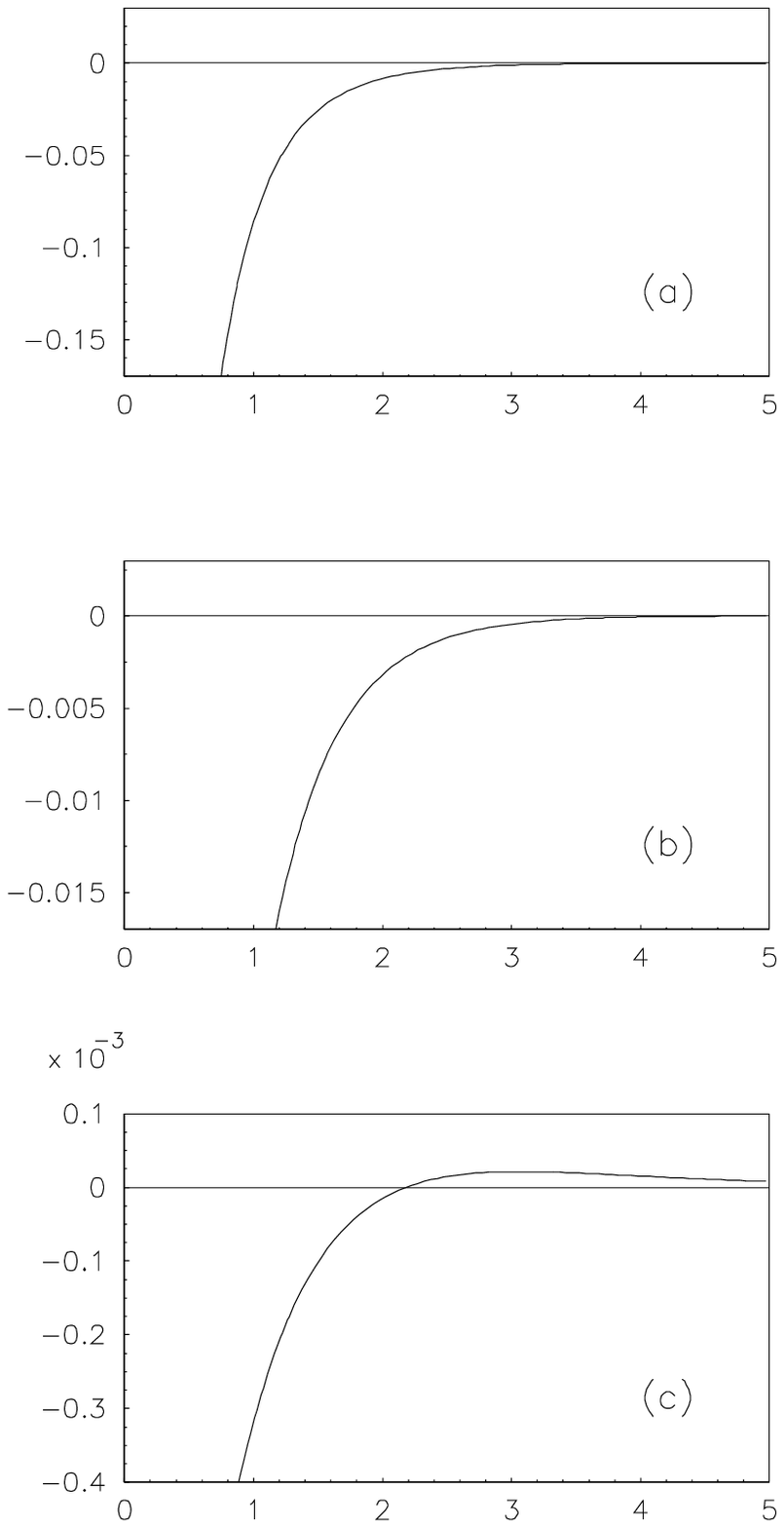,height=19cm,width=10cm}
 \end{minipage}
 \vspace*{-1cm}
\caption{The antirestoring acceleration for the overdamped modes 
$\tilde{y} _{o}$ at the 
values of the $\gamma$ parameter: (a) 1; (b) 1/2; (c) 1/10.}
 \vspace*{-3mm}
\label{fig6}
\end{figure}



\begin{thebibliography} {99}



\bibitem{do}
        W. Hauser, {\em Introduction to the Principles of Mechanics},
        (Addison-Wesley, 1965) pp. 106-113; G.R. Fowles, {\em Analytical
        Mechanics} (CBS College Publishing, 1986) pp. 64-68;
        J.D. Garrison, Am. J. Phys. {\bf 42}, 694 (1974);
        {\em ibid} {\bf 43}, 463 (1975).

\bibitem{W}
        E. Witten,
        Nucl. Phys. B {\bf 185}, 513 (1981).

\bibitem{M}
     B. Mielnik,
     J. Math. Phys. {\bf 25}, 3387 (1984).

\bibitem{D}
     G. Darboux,
     C.R. Acad. Sci. {\bf 94}, 1456 (1882).

\bibitem{tb}
     D.J. Inman, {\em Engineering Vibration} (Prentice Hall, Englewood Cliffs,
     NJ, 1996)


\end{thebibliography}
\end{document}